\documentclass[showpacs,showkeys,pre,footinbib]{revtex4-1}
\usepackage{amsmath}
\usepackage{hyperref}
\usepackage{graphicx}

\usepackage{natbib} 			
\usepackage{amsfonts}
\usepackage{amsmath}
\usepackage{amssymb}
 \usepackage{amsthm}
\usepackage{booktabs}
\usepackage{epsfig}

\begin{document}
\title{Spectrum-based estimators of the bivariate Hurst exponent}
\author{Ladislav Kristoufek\footnote{Institute of Information Theory and Automation, Academy of Sciences of the Czech Republic, CZ-182 08, E-mail: kristouf@utia.cas.cz}\footnote{Institute of Economic Studies, Faculty of Social Sciences, Charles University in Prague, Czech Republic, CZ-110 00}}

\begin{abstract}
We introduce two new estimators of the bivariate Hurst exponent in the power-law cross-correlations setting -- the cross-periodogram and local $X$-Whittle estimators -- as generalizations of their univariate counterparts. As the spectrum-based estimators are dependent on a part of the spectrum taken into consideration during estimation, a simulation study showing performance of the estimators under varying bandwidth parameter as well as correlation between processes and their specification is provided as well. The newly introduced estimators are less biased than the already existent averaged periodogram estimator which, however, has slightly lower variance. The spectrum-based estimators can serve as a good complement to the popular time domain estimators. 
\end{abstract}

\pacs{05.10.-a, 05.45.-a, 89.65.Gh}
\keywords{power-law cross-correlations, long-term memory, spectrum-based estimators}

\maketitle

\section{Introduction}

Introduction of the detrended cross-correlation analysis (DCCA) \cite{Podobnik2008,Zhou2008,Jiang2011,Podobnik2011} has opened a new branch of studying cross-correlations between two series which has found its home in various disciplines -- hydrology and (hydro)meteorology \cite{Hajian2010,Vassoler2012,Kang2013}, seismology and geophysics \cite{Shadkhoo2009,Marinho2013}, biology and biometrics \cite{Xue2012, Ursilean2009}, DNA sequences \cite{Stan2013}, neuroscience \cite{Jun2012}, music \cite{Hennig2014}, electricity \cite{Wang2013b}, finance \cite{Podobnik2009a,Lin2012,Shi2013}, commodities \cite{He2011,He2011a}, traffic \cite{Zebende2009,Xu2010,Zhao2011} and others. Since then, height cross-correlation analysis (HXA) \cite{Kristoufek2011} and detrending moving-average cross-correlation analysis (DMCA) \cite{Arianos2009,He2011a} have been introduced as complements to DCCA. These estimators are based on a generalization of the power-law decay of the autocorrelation function $\rho(k)$ with lag $k$ in the univariate case where $\rho(k) \propto k^{2H-2}$ for lag $k \rightarrow +\infty$. Hurst exponent $H$ is a measure of long-range dependence and it holds that $0 \le H < 1$ for stationary series. For $H>0.5$, the process is positively long-range correlated and reminds of a trending process which, however, returns back to its mean and remains stationary. Due to this characteristic, such processes are also usually referred to as persistent processes. From the other side, $H<0.5$ implies anti-persistence which features frequent reversal of signs and in general negative correlation. No long-range dependence is present for $H=0.5$.

Long-range correlations are generalized into the bivariate setting simply by translating the properties into two dimensions -- a power-law decay of the cross-correlation function of the two analyzed processes. Specifically, we have the cross-correlation function $\rho_{xy}(k)$ with lag $k$ of a specific form, i.e. $\rho_{xy}(k) \propto k^{2H_{xy}-2}$ for lag $k \rightarrow +\infty$. If the bivariate Hurst exponent $H_{xy}>0.5$, we have power-law cross-correlated processes, or alternatively cross-persistent or long-range cross-correlated processes. Such processes are persistent in their co-movement, i.e. if the pair of series moved together in the past, they are more likely to move together in the future as well \cite{Kristoufek2011}. Such definition is positioned in the time domain. However, we can also approach the power-law cross-correlations in the frequency domain.

An alternative definition of the power-law cross-correlated processes is again built on a generalization of the univariate definition. The cross-persistent process is characterized by the cross-power spectrum of a form $|f_{xy}(\lambda)|\propto \lambda^{1-2H_{xy}}$ for frequency $\lambda \rightarrow 0+$ \cite{Sela2012}. The power-law asymptotic decay of the cross-correlation function thus simply translates to the power-law divergence of the cross-power spectrum close to the origin. This alternative approach to the power-law cross-correlations is used only marginally in the interdisciplinary physics literature. 

Here, we introduce two new estimators of the bivariate Hurst exponent based on the spectral definition of the power-law cross-correlations. Specifically, we establish the cross-periodogram estimator and the local $X$-Whittle estimator as generalizations of their univariate counterparts, we discuss the properties of these estimators and we provide a comparison to the only existent spectrum-based estimator -- the averaged periodogram estimator \cite{Sela2012}. In the next section, the estimators are described in detail. The following section discusses the accuracy of the estimators with a changing bandwidth parameter. And the last section concludes.

\section{Spectrum-based estimators}
Frequency domain estimators are based on the definition that at frequencies close to the origin, the magnitude of the cross-power spectrum follows a power law and diverges to infinity. Estimation of the cross-power spectrum thus becomes crucial. The most frequently used tool is a cross-periodogram $I_{xy}(\lambda)$ defined as
\begin{equation}
\label{eq:cross-periodogram}
I_{xy}(\lambda_j)=\frac{1}{2\pi}\sum_{k=-\infty}^{+\infty}{\widehat{\gamma}_{xy}(k)\exp(-i\lambda_jk)}=\frac{1}{2\pi T}\sum_{t=1}^{T}{x_t\exp(-i\lambda_j t)}\sum_{t=1}^{T}{y_t\exp(i\lambda_j t)}=I_x(\lambda_j)\overline{I_y(\lambda_j)},
\end{equation}
where $T$ is the time series length, $\widehat{\gamma}_{xy}(k)$ is an estimated cross-covariance at lag $k$ and $\lambda_j$ is a frequency defined as $\lambda_j=2\pi j/T$ where $j=1,2,\ldots,\lfloor T/2 \rfloor$ and $\lfloor \rfloor$ is the nearest lower integer operator so that the cross-periodogram is defined between 0 and $\pi$ only. $I_x(\lambda_j)$ is a periodogram of series $\{x_t\}$ and $\overline{I_y(\lambda_j)}$ is a complex conjugate of a periodogram of series $\{y_t\}$. If the cross-periodogram is used in its raw form, then evidently, we automatically obtain $H_{xy}=\frac{H_x+H_y}{2}$. Moreover, the raw cross-periodogram (as well as the raw univariate periodogram) is an inconsistent estimator of the true cross-power spectrum \citep{Wei2006}. To overcome the inconsistency issue, the raw (cross-)periodogram needs to be smoothed. Bloomfield \cite{Bloomfield2000} suggests a simple smoothing operator based on Daniell \cite{Daniell1946}, which is practically a simple moving average with half weights on the boundary values. Some authors \citep{Velasco1999,Hurvich2000,Sela2012} also suggest to first taper the series to deal with leakages at low frequencies. In the following text, we apply only the smoothing of periodograms while the tapering is not utilized as it did not show any finite sample efficiency or bias gains for the estimators we use. In this section, we cover three estimators of the bivariate Hurst exponent $H_{xy}$, two of which are newly introduced here.

\subsection{Averaged periodogram estimator}

Sela \& Hurvich \cite{Sela2012} propose the averaged periodogram estimator (APE) and they are in fact the first ones to propose an estimator of $H_{xy}$ (or more precisely $d_{12}$ in their case, where  it holds that $d_{12}=H_{xy}-0.5$ as in the univariate case) in the frequency domain. The estimator is a bivariate generalization of the method of Robinson \cite{Robinson1994}. Taking the cumulative cross-periodogram $\widehat{F_{xy}}(\lambda)=\frac{2\pi}{m}\sum_{j=1}^{\lfloor m\lambda/2\pi \rfloor}{I_{xy}(\lambda_j)}$ where $m\le T/2$ is a bandwidth parameter and fixed $q\in (0,1)$, the estimator is given by

\begin{equation}
\widehat{H_{xy}}=1-\frac{\log\frac{\widehat{F_{xy}}(q\lambda_m)}{\widehat{F_{xy}}(\lambda_m)}}{2\log q}.
\end{equation}

Under twelve assumptions given by Sela \& Hurvich \cite{Sela2012}, the estimator is consistent. Moreover, it is advised to use $q=0.5$. The authors also provide a Monte Carlo simulation study to show the finite sample properties of the estimator. The bias and efficiency are shown by box plots for several scenarios to show that for samples below 10,000 observations, the estimator is strongly biased with high variance. For high values of $m$ and high number of observations, the variance of the estimator decreases markedly while the estimator still remains biased. 

\subsection{Cross-periodogram estimator}

As given in the definition of the cross-persistent processes, the cross-power spectrum diverges at the origin as a power law with exponent $1-2H_{xy}$. Using the cross-periodogram as an estimator of the cross-power spectrum, we expect the long-range cross-correlated series to follow
\begin{equation}
\label{eq:CP_H}
|I_{xy}(\lambda_j)| \propto \lambda_j^{1-2H_{xy}}.
\end{equation}
The cross-periodogram estimator (XPE) of the bivariate Hurst exponent $H_{xy}$ can be obtained through a regression on
\begin{equation}
\label{eq:CP_H_reg}
\log|I_{xy}(\lambda_j)| \propto -(2H_{xy}-1)\log\lambda_j.
\end{equation}

As the power-law scaling is expected only for $\lambda \rightarrow 0+$, the regression is not performed over all frequencies. By choosing $\lambda_j=2\pi j/T$ for $j=1,2,\ldots,m$ where $m\le T/2$, we estimate the bivariate Hurst exponent using only the information up to a selected frequency based on a selection of the bandwidth parameter $m$. For the univariate case, Beran \cite{Beran1994} and Robinson \cite{Robinson1995} show that the periodogram estimator is consistent and asymptotically normal with 
\begin{equation}
\sqrt{m}(\widehat{H}-H^0) \rightarrow_d N(0,\pi^2/24)
\end{equation}
where $H^0$ is the true Hurst exponent. The limiting distribution is free of $H^0$ and also of all the other parameters (the assumptions are given in Theorem 4.6 of Beran \cite{Beran1994}). The variance of the estimator decreases with a square root of $m$ but we need to keep in mind that the higher the $m$ parameter is, the more biased the estimator is because the power-law scaling holds only for the origin neighborhood. Choice of $m$ thus depends on preferences between bias and efficiency. For the bivariate case, however, we have more parameters in the specification of the model -- mainly the univariate Hurst exponents of the separate processes and the correlation coefficient between error-terms for the simplest case -- and there is no reason to believe that the properties of the XPE estimator would be independent of these. Showing the asymptotic properties requires a strict set of assumptions and the underlying bivariate model specification. In the context of this text, it would be out of line to assume some particular specification of the underlying model as we try to keep the assumptions of the methods as general as possible. We thus do not provide the asymptotic properties for this estimator and leave it for further research, yet still, we provide a discussion about dependence of mean and variance of the estimator on the set of parameters and compare the properties with the other two frequency-based estimators later in the next section.

\subsection{Local $X$-Whittle estimator}

The local Whittle estimator of the fractional differencing parameter $d$ or Hurst exponent $H$ is based on the same principle as the previously defined periodogram estimator -- the power-law divergence of the power spectrum. However, instead of the regression fitting to the power-law scaling near the origin as $\lambda\rightarrow 0+$, the local Whittle estimator is based on a minimization of the penalty function based on K\"unsch \cite{Kunsch1987}. 

Taking the work of Robinson \cite{Robinson1995a} as a starting point and generalizing the method for the bivariate series, we propose the estimator of the bivariate Hurst exponent $H_{xy}$ as follows. Divergence of the magnitude of the cross-power spectrum close to the origin with the power-law scaling is assumed for long-range cross-correlated processes $\{x_t\}$ and $\{y_t\}$. Cross-periodogram $I_{xy}(\lambda)$ is defined according to Eq. \ref{eq:cross-periodogram} with $j=1,2,\ldots,m$ where $m\le T/2$ and $\lambda_j=2\pi j/T$. Assuming that series $\{x_t\}$ and $\{y_t\}$ are indeed long-range cross-correlated with $\frac{1}{2}<H_{xy}\le 1$, we propose the local $X$-Whittle estimator (LXW) as
\begin{equation}
\label{eq:LWX}
\widehat{H_{xy}}=\arg \min_{\frac{1}{2}<H_{xy}\le 1} R(H_{xy}),
\end{equation} 
where 
\begin{equation}
\label{eq:LWX_R}
R(H_{xy})=\log\left(\frac{1}{m}\sum_{j=1}^m{\lambda_j^{2H_{xy}-1}|I_{xy}(\lambda_j)|}\right)-\frac{2H_{xy}-1}{m}\sum_{j=1}^m{\log \lambda_j}
\end{equation}
and $\lambda_j=2\pi j/T$. Eq. \ref{eq:LWX_R} represents the likelihood function of K\"{u}nsch \cite{Kunsch1987}. The LXW estimator is thus a semi-parametric maximum likelihood estimator as it utilizes only the properties of the cross-power spectrum near the origin.

In a similar manner as for the XPE estimator, the univariate version of the local Whittle estimator is consistent and asymptotically normal, specifically 
\begin{equation}
\sqrt{m}(\widehat{H}-H^0) \rightarrow_d N(0,1/4)
\end{equation}
where again $H^0$ is the true bivariate Hurst exponent and the limiting distribution is free of $H^0$ and all the other parameters. The local Whittle estimator is thus asymptotically more efficient than the periodogram estimator as $1/4<\pi^2/24$. For detailed treatment of the univariate case and the assumptions, see \cite{Robinson1995a}. In the bivariate case, there is again no reason to presume that the asymptotic properties would be independent of the univariate Hurst exponents and the correlation structure of the error-terms. Discussion of these possible dependencies and comparison with APE and XPE are provided in the following section \footnote{R-scripts of all three estimators are available upon request from the author.}. 

\section{Dependence on bandwidth parameter}

Choice of the bandwidth parameter $m$ is a crucial aspect of the frequency domain estimators as hinted in the previous section. As noted in the studies dealing with the univariate specifications of the estimators \citep{Beran1994,Robinson1995a,Robinson1995}, variance should decrease with the parameter and bias should increase. The former comes from the fact that the estimation is based on more data points and latter from the fact that the power-law scaling of the cross-power spectrum holds only for the lowest frequencies. Here, we present and discuss the behavior of the mean and variance of the frequency-based estimators presented above with respect to varying parameter $m$.

We discuss two main scenarios of the long-range cross-correlated processes -- the processes with and without power law coherency behavior, i.e. when the bivariate Hurst exponent $H_{xy}$ is not or is, respectively, equal to the average of the separate Hurst exponents $H_x$ and $H_y$. For the latter (simpler) case, we utilize ARFIMA(0,$d$,0) -- autoregressive fractionally integrated moving average -- processes with correlated error-terms and with $d_1=d_2=0.4$, and for the former, we use the mixed-correlated ARFIMA processes \cite{Kristoufek2013} with $d_1=d_4=0.4$ and $d_2=d_3=0.2$. 

Specifically, for given $d$ parameters and $a_n(d)=\frac{\Gamma(n+d)}{\Gamma(n+1)\Gamma(d)}$, the correlated ARFIMA processes $\{x_t\}$ and $\{y_t\}$ are given as
\begin{gather}
\label{eq1}
x_t=\sum_{n=0}^{\infty}{a_n(d_1)\varepsilon_{t-n}} \\
y_t=\sum_{n=0}^{\infty}{a_n(d_2)\nu_{t-n}} \nonumber \\
\langle \varepsilon_t \rangle = \langle \nu_t \rangle = 0 \nonumber\\
\langle \varepsilon_t^2 \rangle = \sigma_{\varepsilon}^2 < +\infty \nonumber\\
\langle \nu_t^2 \rangle = \sigma_{\nu}^2 <+\infty \nonumber\\
\langle \varepsilon_t\varepsilon_{t-n} \rangle = \langle \nu_t\nu_{t-n} \rangle = \langle \varepsilon_t\nu_{t-n} \rangle = 0\text{ for }n \ne 0 \nonumber\\
\langle \varepsilon_t\nu_t \rangle = \sigma_{\varepsilon\nu} <+\infty \nonumber.
\label{eq:ARFIMA_varcovar}
\end{gather}
The mixed-correlated ARFIMA processes are defined as
\begin{gather}
x_t=\sum_{n=0}^{+\infty}{a_n(d_1)\varepsilon_{1,t-n}}+\sum_{n=0}^{+\infty}{a_n(d_2)\varepsilon_{2,t-n}} \\
y_t=\sum_{n=0}^{+\infty}{a_n(d_3)\varepsilon_{3,t-n}}+\sum_{n=0}^{+\infty}{a_n(d_4)\varepsilon_{4,t-n}} \nonumber\\
\langle \varepsilon_{i,t} \rangle = 0\text{ for }i=1,2,3,4 \nonumber\\
\langle \varepsilon_{i,t}^2 \rangle = \sigma_{\varepsilon_i}^2\text{ for }i=1,2,3,4 \nonumber\\
 \langle \varepsilon_{i,t}\varepsilon_{j,t-n} \rangle = 0\text{ for }n \ne 0\text{ and }i,j=1,2,3,4 \nonumber\\
\langle \varepsilon_{i,t}\varepsilon_{j,t} \rangle = \sigma_{ij}\text{ for }i,j=1,2,3,4\text{ and }i\ne j. \nonumber
\label{eq:ARFIMA_LC}
\end{gather}

Both kinds of processes are studied for the time series with $T=5000$ observations and correlation between error-terms varying in an interval between 0.2 and 1 with a step of 0.2. To uncover the dependence on $m$, we use $m/T$ from 0.05 up to 0.5 with a step of 0.05. We thus cover the cross-periodogram from the lowest tenth of the frequencies up to the whole cross-periodogram. For each specification, we use 1,000 simulations and the Daniell's window of 21 as used in Sela \& Hurvich \cite{Sela2012}. This way, we are able to comment on the dependence of bias and variance of the estimator with respect to the correlation between error-terms and the bandwidth parameter $m$.

\begin{figure}[!htbp]
\begin{center}
\begin{tabular}{cc}
\includegraphics[width=83mm]{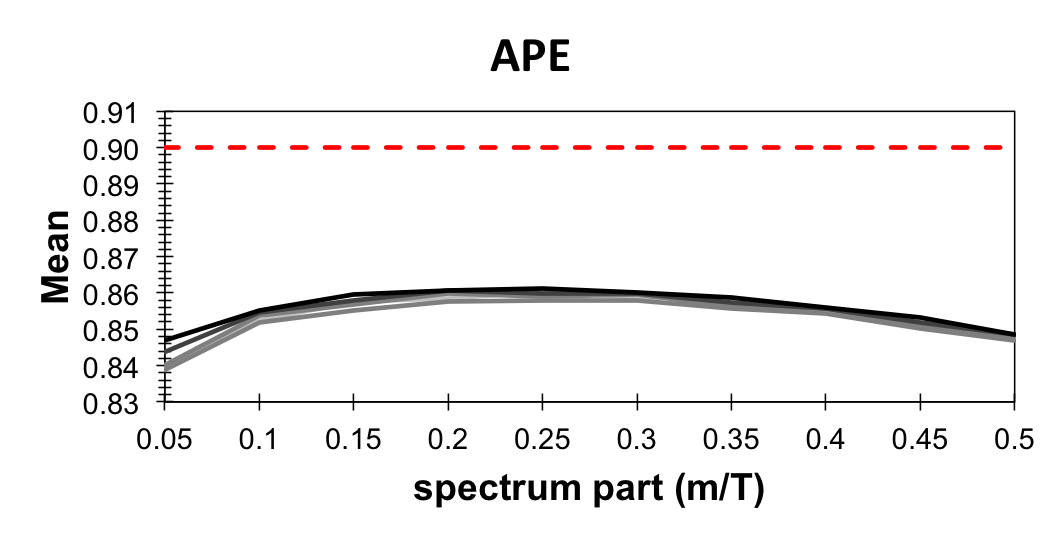}&\includegraphics[width=83mm]{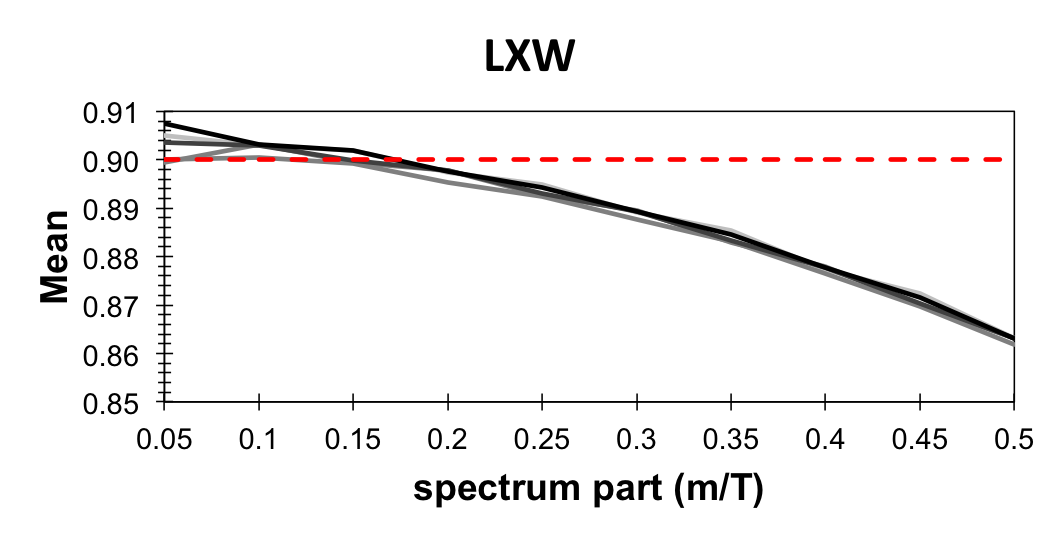}
\end{tabular}
\begin{tabular}{c}
\includegraphics[width=83mm]{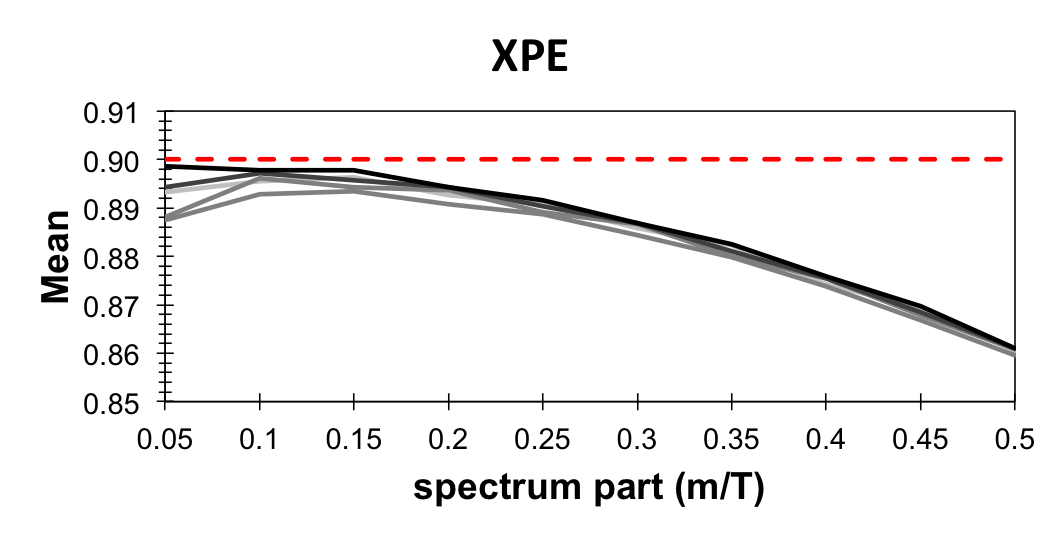}
\end{tabular}
\end{center}\vspace{-0.5cm}
\caption[Mean values of APE, LXW and XPE estimators I]{\textbf{(Color online) Mean values of APE, LXW and XPE estimators dependent on $m$ and correlation between error-terms I.} \footnotesize{Values are based on 1,000 simulations of ARFIMA(0,$d$,0) processes with correlated error-terms and $d_1=d_2=0.4$. Correlation between error-terms ranges between 0.2 and 1 with a step of 0.2 and is represented by different shades of grey in the chart (the lightest for 0.2 and black for 1). Red line represents the true value of $H_{xy}=0.9$.}\label{fig:m_1}
}
\end{figure}

\begin{figure}[!htbp]
\begin{center}
\begin{tabular}{cc}
\includegraphics[width=83mm]{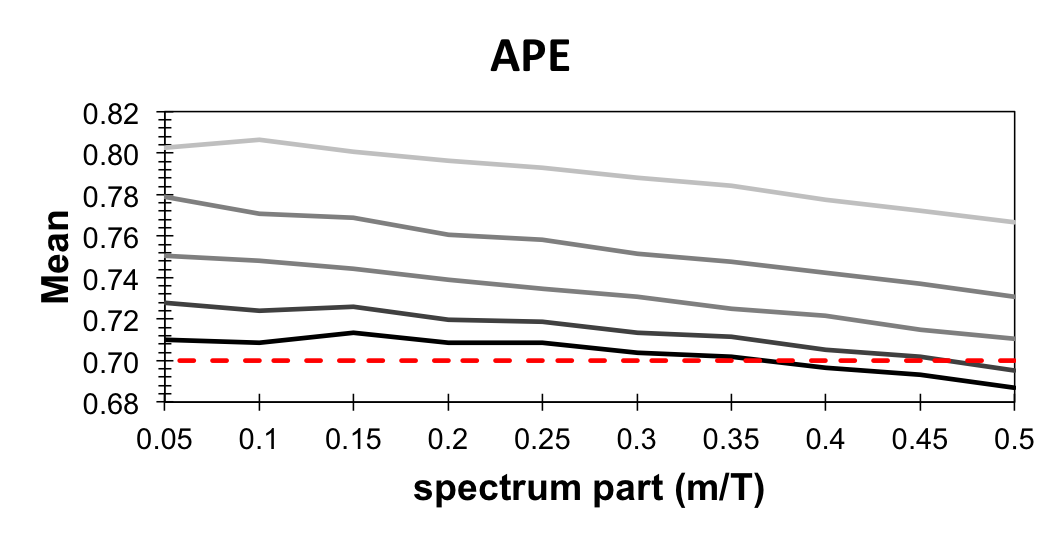}&\includegraphics[width=83mm]{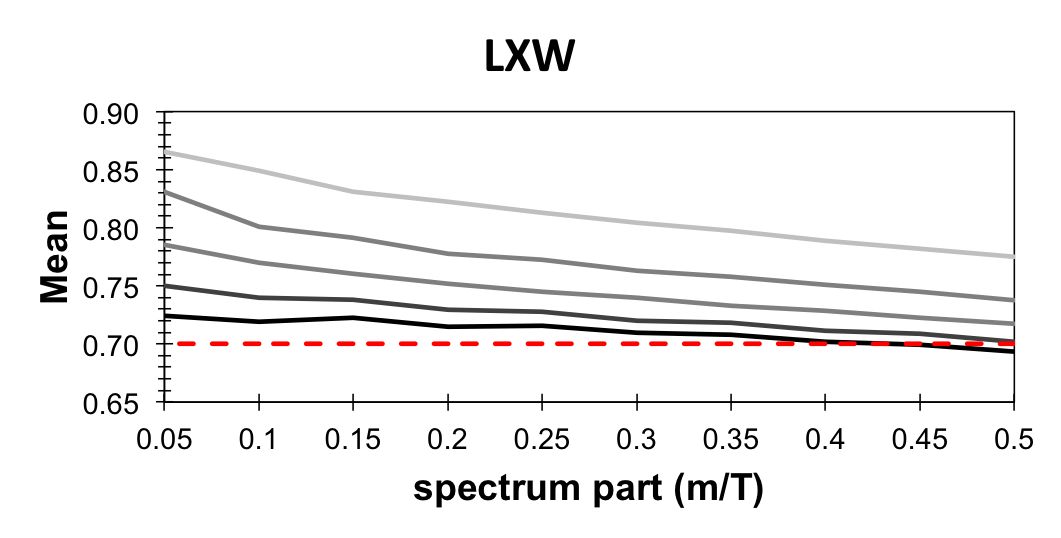}
\end{tabular}
\begin{tabular}{c}
\includegraphics[width=83mm]{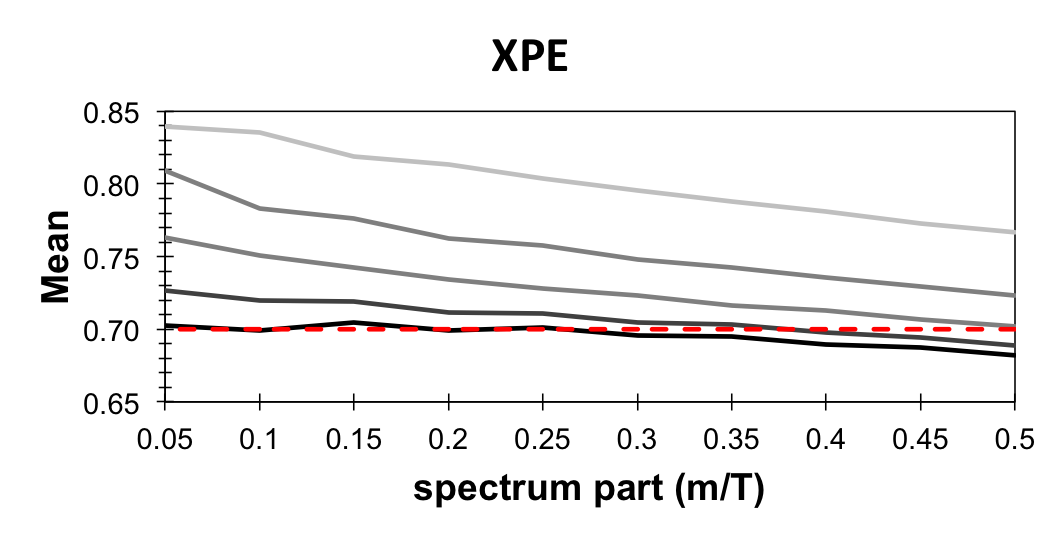}
\end{tabular}
\end{center}\vspace{-0.5cm}
\caption[Mean values of APE, LXW and XPE estimators II]{\textbf{(Color online) Mean values of APE, LXW and XPE estimators dependent on $m$ and correlation between error-terms II.} \footnotesize{Values are based on 1,000 simulations of mixed-correlated ARFIMA(0,$d$,0) processes with $d_1=d_4=0.4$ and $d_2=d_3=0.2$. Correlation between error-terms ranges between 0.2 and 1 with a step of 0.2 and is represented by different shades of grey in the chart (the lightest for 0.2 and black for 1). Red line represents the true value of $H_{xy}=0.7$.}\label{fig:m_3}
}
\end{figure}

Starting with the bias, Figs. \ref{fig:m_1} and \ref{fig:m_3} show the results of simulations for the correlated and mixed-correlated ARFIMA processes. In Fig. \ref{fig:m_1}, we observe that for the correlated ARFIMA processes for which we expect the bivariate Hurst exponent to be equal to the average of the separate Hurst exponents \cite{Podobnik2009,Kristoufek2013}, the bias behavior differs for specific methods. For LXW and XPE, we observe an expected behavior -- the estimates are unbiased for approximately $m/T\le 0.2$, i.e. the lowest frequencies. For higher values of $m$, the estimates become biased downwards. Interestingly, the bias is practically independent of the correlation level between error-terms for all three estimators. However, for APE, the mean values of the estimates are very stable across various $m$ but remain well below the theoretical value of 0.9 and yield a negative bias of approximately $-0.05$. Again, the bias is practically independent of the correlations between error-terms. 

The situation is more interesting for the mixed-correlated ARFIMA processes, i.e. the power law coherency case. In Fig. \ref{fig:m_3}, we can see that the mean values of the estimates are dependent on both $m$ and the correlation between error-terms for all three estimators. In general, it holds that the estimates are less biased with an increasing correlation between error-terms, which is expected, but also with higher $m$. The performance of the estimators is thus not only dependent on the parameters but also on the specification of the model as shown by the difference between the two cases.

\begin{figure}[!htbp]
\begin{center}
\begin{tabular}{cc}
\includegraphics[width=83mm]{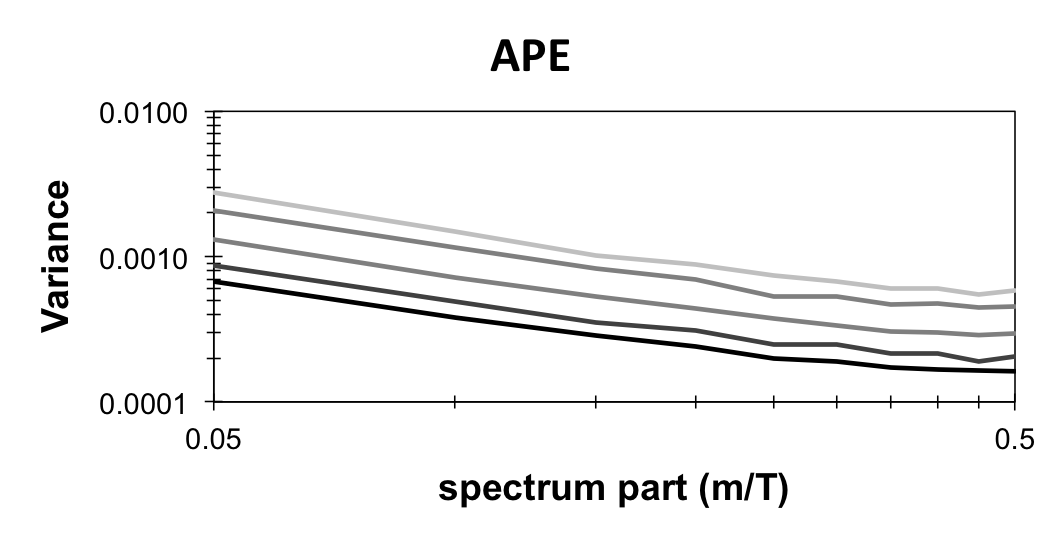}&\includegraphics[width=83mm]{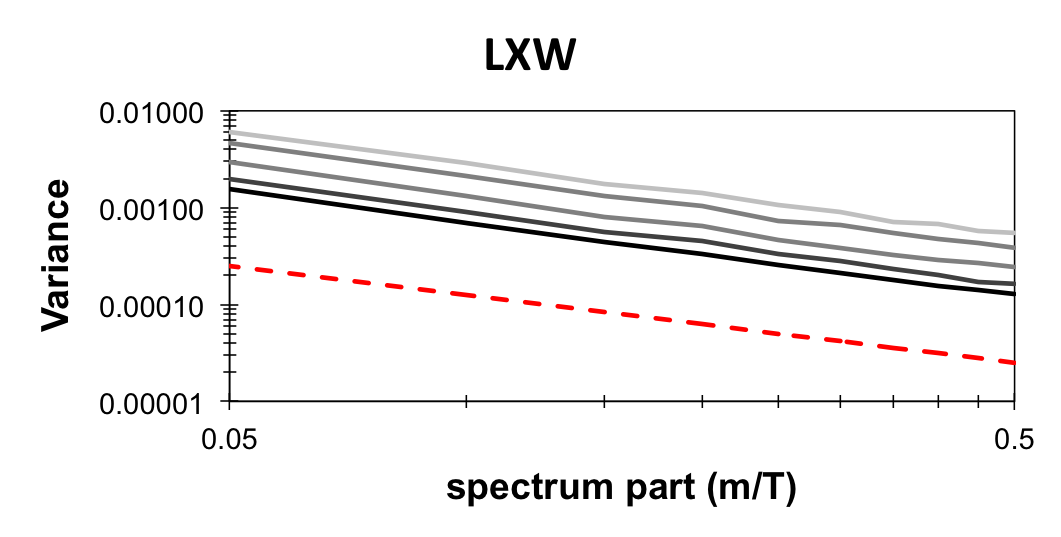}
\end{tabular}
\begin{tabular}{c}
\includegraphics[width=83mm]{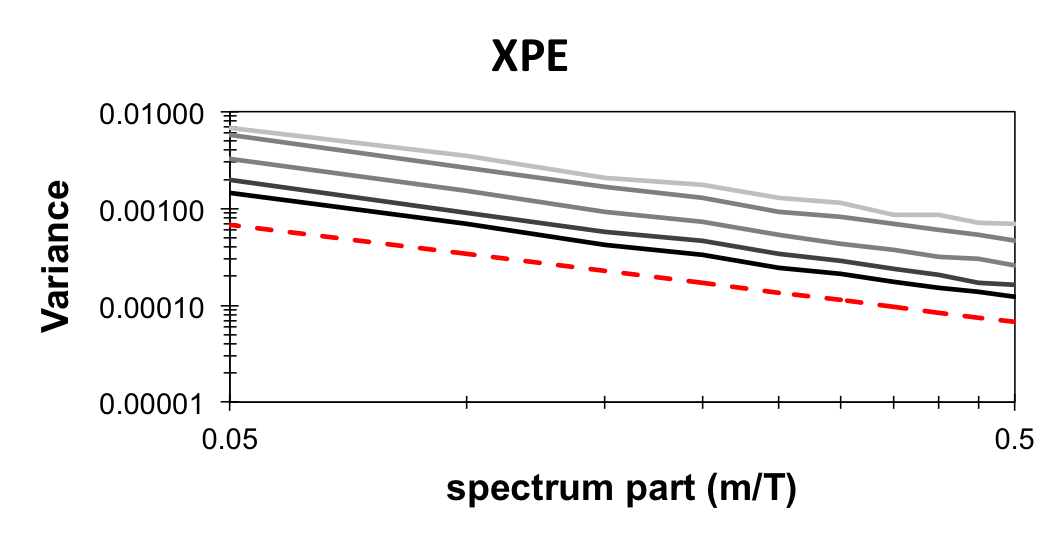}
\end{tabular}
\end{center}\vspace{-0.5cm}
\caption[Variance of APE, LXW and XPE estimators I]{\textbf{(Color online) Variance of APE, LXW and XPE estimators dependent on $m$ and correlation between error-terms I.} \footnotesize{Values are based on 1,000 simulations of ARFIMA(0,$d$,0) processes with correlated error-terms and $d_1=d_2=0.4$. Correlation between error-terms ranges between 0.2 and 1 with a step of 0.2 and is represented by different shades of grey in the chart (the lightest for 0.2 and black for 1). Red line represents the asymptotical values for the univariate case (for LXW and XPE only).}\label{fig:m_2}
}
\end{figure}

The situation is quite similar for the behavior of variance of the estimators. In Fig. \ref{fig:m_2}, we show this behavior compared to the theoretical asymptotic variance for the univariate cases (for LXW and XPE). We observe several regularities which are true for all three estimators. First, the variance decreases with the increasing $m$ as expected. Second, the variance decreases with the increasing strength of correlations between error-terms. Third, the log-log depiction indicates a power-law scaling with the parameter $m$. For LXW and XPE, we can compare this scaling with the asymptotic scaling for the respective univariate estimators and it is visible that these can be seen as square-root scalings which are in hand with the univariate case. However, the levels of variance are well above the asymptotic univariate values. The variance levels of LXW and XPE practically overlap while the variance of APE shows slightly lower values.

\begin{figure}[!htbp]
\begin{center}
\begin{tabular}{cc}
\includegraphics[width=83mm]{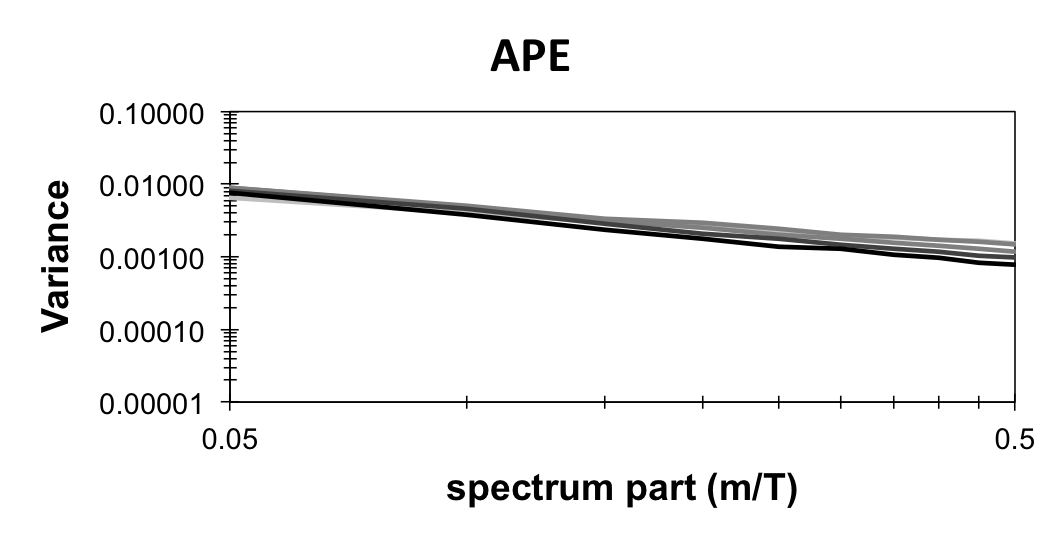}&\includegraphics[width=83mm]{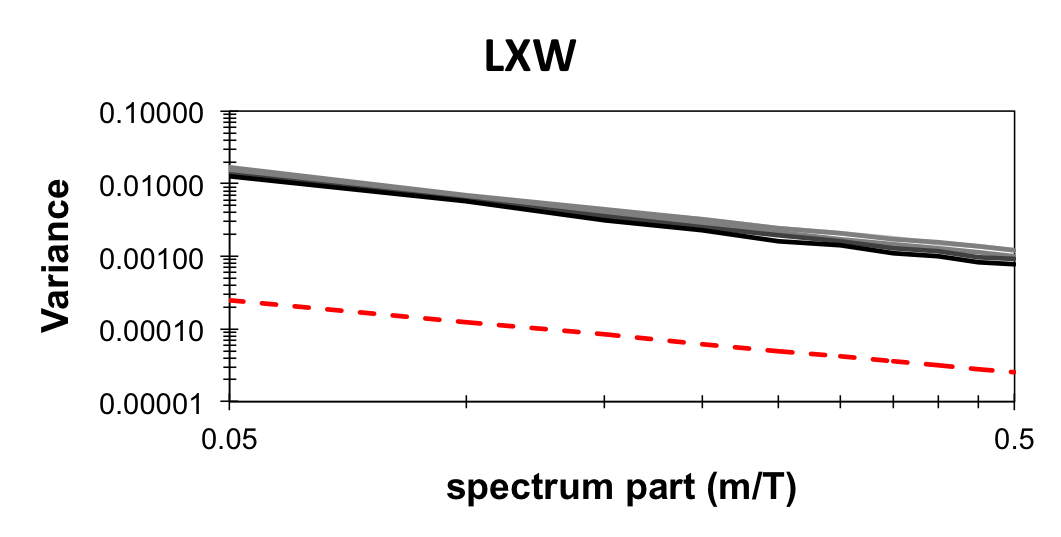}
\end{tabular}
\begin{tabular}{c}
\includegraphics[width=83mm]{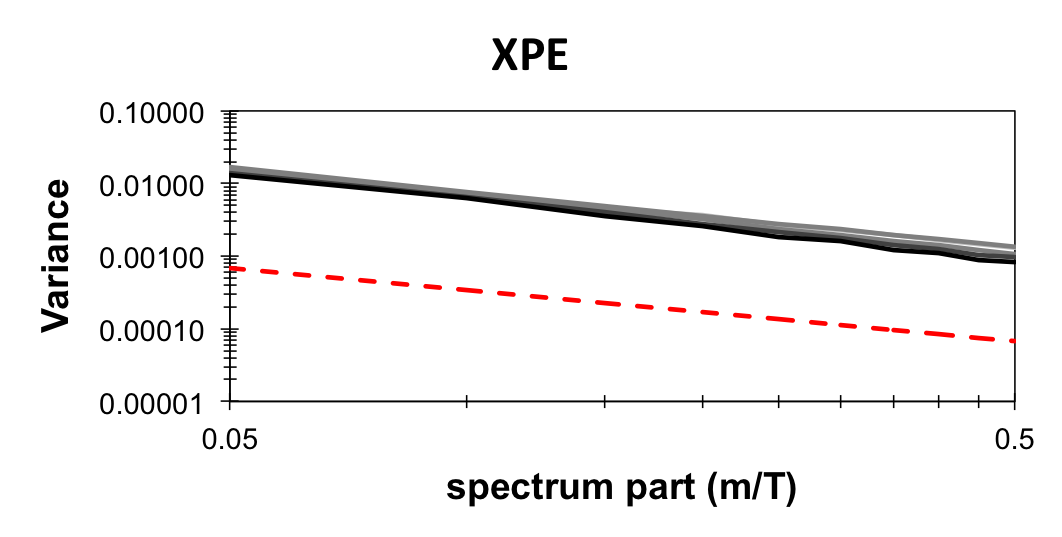}
\end{tabular}
\end{center}\vspace{-0.5cm}
\caption[Variance of APE, LXW and XPE estimators II]{\textbf{(Color online) Variance of APE, LXW and XPE estimators dependent on $m$ and correlation between error-terms II.} \footnotesize{Values are based on 1,000 simulations of mixed-correlated ARFIMA(0,$d$,0) processes with $d_1=d_4=0.4$ and $d_2=d_3=0.2$. Correlation between error-terms ranges between 0.2 and 1 with a step of 0.2 and is represented by different shades of grey in the chart (the lightest for 0.2 and black for 1).}\label{fig:m_4}
}
\end{figure}

For the mixed-correlated ARFIMA processes, Fig. \ref{fig:m_4} depicts the behavior of variance. All the regularities from the previous paragraph hold even here. However, the level of variances is evidently much higher compared to the previous case and it shows that the estimators have much higher variance for the power law coherency specification. Even though the variance level again depends on the level of correlation between error-terms, the general variance level dominates. As a result, the distance from the univariate asymptotic variances of LXW and XPE is much more profound as well. Again, the level of variances is lower for APE compared to the other two estimators. 

\section{Conclusions}

We introduce two new estimators of the bivariate Hurst exponent -- the cross-periodogram and local $X$-Whittle estimators -- in the power-law cross-correlations setting and compare them with the already existent averaged periodogram estimator. As the spectrum-based estimators depend on a part of the spectrum taken into consideration during estimation, in the same way as the time domain estimators depend on the utilized scales, we also provide a simulation study showing the performance of the three estimators under varying bandwidth parameter as well as correlation between processes and their specification. The newly introduced estimators are less biased than the averaged periodogram which, however, has slightly lower variance. The spectrum-based estimators can serve as a good complement to the already existing and popular time domain estimators. 

\section*{Acknowledgements}
Support from the Czech Science Foundation under project No. 14-11402P and the Grant Agency of the Charles University in Prague under project No. 1110213 is gratefully acknowledged.

\newpage

\bibliography{Bibliography}
\bibliographystyle{unsrt}

\newpage

\end{document}